\documentclass[prx,twocolumn,letterpaper,superscriptaddress]{revtex4}
\usepackage{graphicx,amsmath,amssymb,amsfonts,latexsym,color,dcolumn,bm,epsfig,mathtools}
\usepackage[latin1]{inputenc}
\usepackage{enumerate,color,subfigure,multirow,colortbl,color,soul,footmisc}

\def\bbm[#1]{\mbox{\boldmath $#1$}}

\graphicspath{{./Figures/}}

\begin{document}

\title{Giant Casimir torque between rotated gratings and the $\theta=0$ anomaly}

\author{Mauro Antezza}\email{mauro.antezza@umontpellier.fr}
\affiliation{Laboratoire Charles Coulomb (L2C), UMR 5221 CNRS-Universit\'{e} de Montpellier, F- 34095 Montpellier, France}
\affiliation{Institut Universitaire de France, 1 rue Descartes, F-75231 Paris Cedex 05, France}

\author{H.~B. Chan}\email{hochan@ust.hk}
\affiliation{Department of Physics, Center for Metamaterial Research and William Mong Institute of Nano Science and Technology, The Hong Kong University of Science and Technology, Clear Water Bay, Kowloon, Hong Kong, China}

\author{Brahim Guizal}\email{brahim.guizal@umontpellier.fr}
\affiliation{Laboratoire Charles Coulomb (L2C), UMR 5221 CNRS-Universit\'{e} de Montpellier, F- 34095 Montpellier, France}

\author{Valery N. Marachevsky}\email{v.marachevsky@spbu.ru}
\affiliation{St. Petersburg State University, 7/9 Universitetskaya nab., St. Petersburg 199034, Russia}

\author{Riccardo Messina}\email{riccardo.messina@institutoptique.fr}
\affiliation{Laboratoire Charles Coulomb (L2C), UMR 5221 CNRS-Universit\'{e} de Montpellier, F- 34095 Montpellier, France}
\affiliation{Laboratoire Charles Fabry, UMR 8501, Institut d'Optique, CNRS, Universit\'{e} Paris-Saclay, 2 Avenue Augustin Fresnel, 91127 Palaiseau Cedex, France}

\author{M. Wang}\email{mwangak@connect.ust.hk}
\affiliation{Department of Physics  and William Mong Institute of Nano Science and Technology, The Hong Kong University of Science and Technology, Clear Water Bay, Kowloon, Hong Kong, China}

\date{\today}

\begin{abstract}
We study the Casimir torque between two metallic one-dimensional gratings rotated by an angle $\theta$ with respect to each other. We find that, for infinitely extended gratings, the Casimir energy is anomalously discontinuous at $\theta=0$, due to a critical zero-order geometric transition between a 2D- and a 1D-periodic system. This transition is a peculiarity of the grating geometry and does not exist for intrinsically anisotropic materials. As a remarkable practical consequence, for finite-size gratings, the torque per area can reach extremely large values, increasing without bounds with the size of the system. We show that for finite gratings with only 10 period repetitions, the maximum torque is already 60 times larger than the one predicted in the case of infinite gratings. These findings pave the way to the design of a contactless quantum vacuum torsional spring, with possible relevance to micro- and nano-mechanical devices.
\end{abstract}

\maketitle

\section{Introduction}
Fluctuations of physical quantities are ubiquitous in nature. They provide valuable information on physical systems and give rise to   a large variety of phenomena, ranging from phase transitions in condensed matter to the origin and evolution of the universe. In quantum physics the omnipresent quantum fluctuations of the electromagnetic field are associated with Heisenberg's uncertainty principle and can often cause spectacular macroscopic effects. They are at the origin of a force existing between any pair of polarizable neutral bodies separated by vacuum. This phenomenon, known as Casimir-Lifshitz effect~\cite{CasimirProcKNedAkadWet48,Lifshitz}, was first theoretically discussed by H. Casimir in 1948 in the configuration of two parallel perfectly conducting slabs~\cite{CasimirProcKNedAkadWet48} and later generalized by E.~M. Lifshitz to dielectric slabs~\cite{Lifshitz}. This force has been extensively studied both theoretically (mainly exploring the role of temperature, geometry and material properties~\cite{DLP,APSS,Milton,Buhmann}) and experimentally in several different configurations \cite{CasimirDalvit}. The Casimir force, which is related to the spatial variation of the Casimir energy, $F = -\partial_z E(z)$, plays an important role in the linear motion of small nano/micro systems at sub-micron separations. Recently, much theoretical effort has been dedicated to the study of the force between gratings, i.e. periodically nanostructured surfaces~\cite{LambrechtPRL08,DavidsPRA10,IntravaiaPRA12,LussangePRA12,GueroutPRA13,GrahamPRA14,WagnerPRA14,NotoPRA14,MessinaPRA15,BuhmannIntJModPhys16}. Using a variety of theoretical frameworks, several authors have highlighted the promising possibilities offered by these structures in terms of modulation of the force as a function of the grating geometrical features. Moreover, these predictions have been tested in several experiments involving two gratings, sphere--grating or atom--grating configurations~\cite{ChanPRL08,ChiuPRB10,BaoPRL10,BanishevPRL13,IntravaiaNatComm14,Bender,TangNatPhot17}. 

\begin{figure}[t!]
\includegraphics[width=0.48\textwidth]{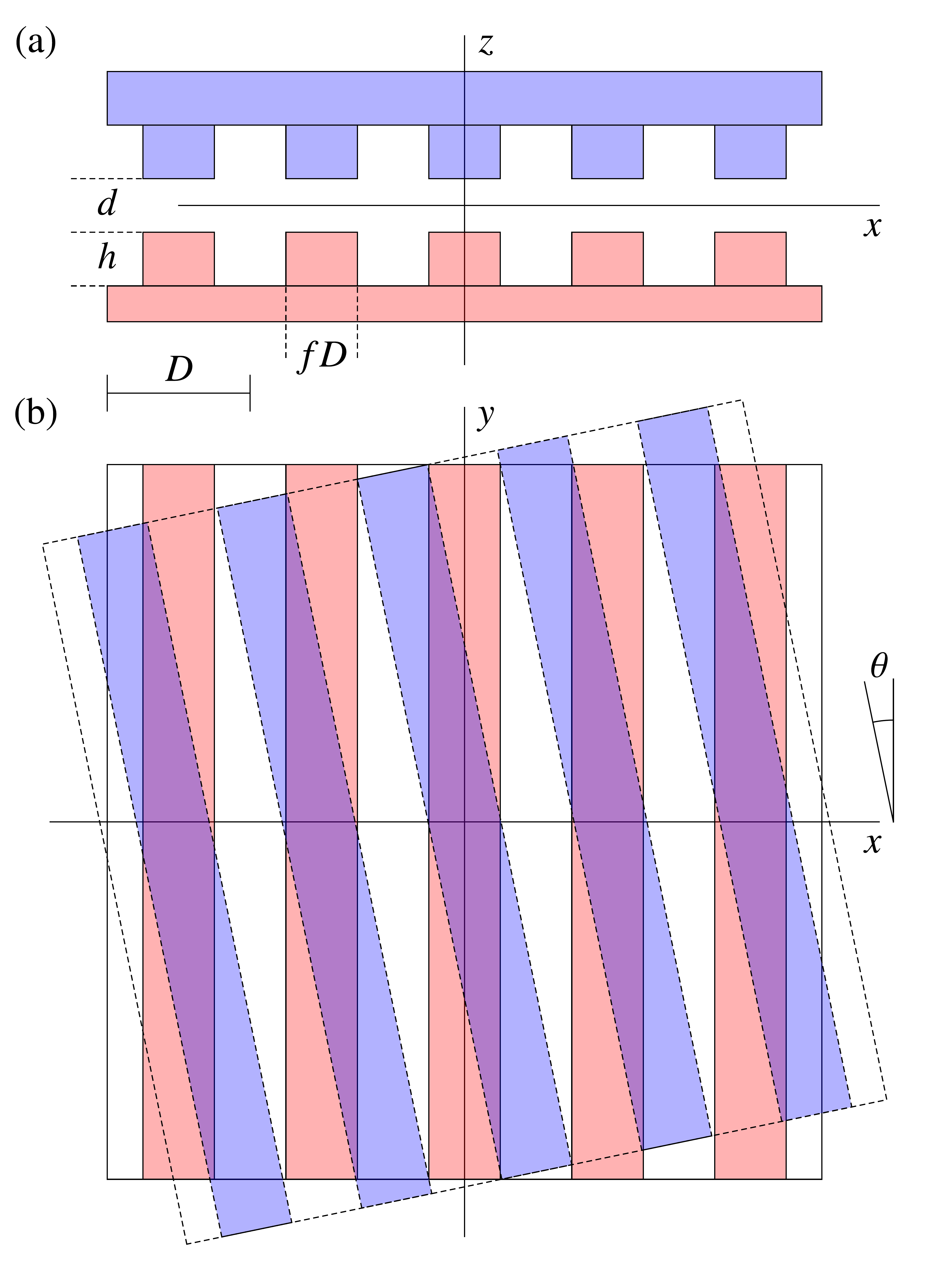}
\caption{Geometry of the system: two one-dimensional periodic lamellar gratings having period $D$ and filling fraction $f$ and height $h$, placed at distance $d$ along the $z$ axis. The figure shows two samples with $n=5$ periods. (a) side view of two aligned gratings. (b) top view of two gratings rotated by an angle $\theta$.}
\label{Geometry}
\end{figure}

In addition to the normal Casimir force, vacuum fluctuations are also exptected to generate a {\it Casimir torque} between closely-spaced anisotropic bodies. Due to the nanostructuration, one important feature of the well-studied grating geometry is the breaking of the rotational symmetry inherent to the configuration of two parallel homogenous infinite flat surfaces. This leads to a further dependence of the Casimir energy of the system $E(z,\theta)$ on the angle $\theta$ between two gratings rotated with respect to the common transversal $z$-axis  [see Fig.~\ref{Geometry}(b)], and thus to a \emph{torque} $\tau = -\partial_\theta E(z,\theta)$ between the two structures, that is absent in the case of two flat homogeneous slabs. The existence and properties of this Casimir torque has been a topic of intense theoretical interest since the seventies ~\cite{Kats71,Parsegian72,Barash78,EnkPRA95,MundayPRA05,RodriguesEPL06,Torres06,CapassoIEEE07,PhilbinPRA08,Siber09,Chen11,Yasui15,XuPRA17,SomersPRL17,GueroutEPL15,Thyiam18,Lindel18,Ali18}, for both gratings and other symmetry-breaking systems. Due to the smallness of the effect, only very recently has the first measurement of this torque been realized~\cite{Somers18}. In these studies, one of the main motivations was the intriguing possibility to manipulate micro- and nano-objects by exploiting rotations induced by the quantum vacuum, in addition to Casimir attraction and repulsion~\cite{Mara,Fial}. 

In this paper our starting point is the study of the Casimir torque between two one-dimensional metallic lamellar gratings, infinitely extending in the directions perpendicular to their separation. Remarkably, and differently from what is commonly accepted and discussed in previous predictions (see for instance \cite{GueroutEPL15}), we show that for infinite gratings the Casimir energy manifests an anomalous discontinuity at $\theta=0$ (i.e. when the gratings are aligned). In particular, we see that the Casimir energy shows a well-defined limiting behavior for $\theta$ going to zero, but this limit does not coincide with the well-known result of two aligned gratings. We test our findings with different numerical approaches, and physically explain this anomaly on the basis of the abrupt geometrical discontinuity of the number of crossing points between the two rotated gratings: there is always an infinite number of them for any finite (however small it may be) non-zero value of the angle $\theta$, while they are entirely absent at $\theta=0$. Based on this first finding we infer that for finite-size gratings (i.e. finite extension in the directions perpendicular to their separation), where the number of crossing points gradually tends to zero for small angles $\theta$, the Casimir energy would tend continuously to its value at $\theta=0$. Nonetheless, the $\theta=0$ anomaly for infinite systems serves as the limiting case for the behavior of the Casimir energy for finite systems, in particular at small angles, where the Casimir energy remains a continuous function (even at $\theta=0$) but tends to mimic the discontinuity observed for infinite gratings via a steeper and steeper behavior close to $\theta=0$ as the system size increases. This implies a \emph{giant} value of the torque per area at small tilting angles between finite-size gratings, increasing \emph{without bounds} as long as the finite-grating size increases. We then confirm these predictions by numerically studying the Casimir energy and torque between finite-size gratings. In particular, we show the existence of a very pronounced peak of the torque for small angles, with an amplitude which rapidly increases with the size of the system. Based on a simple model to determine the smallest angle at which the two gratings cross each other, we give an estimate of the position and of the amplitude of this peak as a function of the system size. In addition, for finite-size gratings we also find oscillations of the Casimir torque, which can remarkably change its sign. We find that the amplitude of these oscillations decreases as the size of the system increases.

\section{Results}

\subsection{Infinite system: the $\theta=0$ anomaly of the Casimir energy}\label{InfSyst}

Let us consider in this section the case of infinitely extending gratings, a system which is strictly periodic. The configuration we consider is made of two identical gold gratings placed at distance $d$ (along the $z$ axis), characterized by their period $D$, their height $h$, and their filling fraction $f$ [$fD$ being the part of the period $D$ filled with material, see Fig.~\ref{Geometry}(a)]. In the following we will consider the distance $d=100\,$nm, and focus on two identical gratings having $D=400\,$nm, $h=200\,$nm and $f=0.5$. The optical properties of gold are described in terms of a Drude model $\varepsilon(\omega)=1-\omega_p^2/\omega(\omega+i\gamma)$, with $\omega_p=9\,$eV and $\gamma=35\,$meV~\cite{Palik98}. Grating 2, on top of grating 1 in Fig.~\ref{Geometry}(b), is rotated by an angle $\theta$ with respect to grating 1. The energy between the two gratings is calculated by exploiting a general theoretical framework for the calculation of the Casimir force and radiative heat transfer between two arbitrary bodies based on the knowledge of their individual scattering operators~\cite{MessinaEurophysLett11,MessinaPRA11}. At thermal equilibrium at a temperature $T$ such that the photon thermal wavelength $\lambda_T=\hbar c/(k_BT)\gg d=100\,$nm (for $T=300K$, $\lambda_T \approx 7.6\mu$m), the purely quantum vacuum fluctuations largely dominate over the thermal fluctuations, and the Casimir energy per unit surface can be calculated using the $T=0$ expression
\begin{equation}\label{EqEn}
 E = \frac{\hbar}{8\pi^3}\int d\xi\int d^2\mathbf{k}\ln\det\Bigl[\mathbb{I}-\mathcal{R}_1(i\xi)e^{-\mathcal{K}d}\mathcal{R}_2(i\xi)e^{-\mathcal{K}d}\Bigr],
\end{equation}
which requires an integration over the imaginary frequencies $\omega=i\xi$ and over the parallel wavevector $\mathbf{k}=(k_x,k_y)$. The properties of the two bodies are taken into account through their reflection operators $\mathcal{R}_1$ and $\mathcal{R}_2$, whereas the distance $d$ separating them appears only in the two exponential factors, where $\mathcal{K}$ is a diagonal operator having elements equal to the $z$ component of the total wavevector $\kappa=\sqrt{\xi^2/c^2+\mathbf{k}^2}$. One key point in the calculation is the choice of a suitable basis with respect to which the integrals in Eq.~\eqref{EqEn} are computed. As suggested in \cite{GueroutEPL15}, it is convenient to rewrite the wavevector $\mathbf{k}$ as 
\begin{equation}\label{kmn}
 \mathbf{k}_{nm} = \mathbf{k} + \frac{2\pi}{D}(n\hat{\mathbf{e}}_x + m\hat{\mathbf{e}}_{x'}),
\end{equation}
where we have introduced the two integer indices $n,m$ and the unit vectors $\hat{\mathbf{e}}_x$ and $\hat{\mathbf{e}}_{x'}$, representing the periodicity axes of the gratings 1 and 2, respectively ($\hat{\mathbf{e}}_{x'}$ thus forms an angle $\theta$ with $\hat{\mathbf{e}}_x$). If $n$ and $m$ span all integer values, it can be easily shown that $\mathbf{k}_{nm}$ covers the entire $(k_x,k_y)$ plane if $\mathbf{k}$ is limited to the first Brillouin zone (FBZ) defined as
\begin{equation}
\begin{dcases}
 -\frac{\pi}{D} \leq k_x \leq \frac{\pi}{D},\\
 -k_x\cot\theta  - \frac{\pi}{D\sin\theta} \leq k_y \leq  -k_x\cot\theta  + \frac{\pi}{D\sin\theta},\\ 
 \Bigl(k_x  - \frac{2\pi}{D}\Bigr)\tan\frac{\theta}{2} \leq k_y \leq   \Bigl(k_x + \frac{2\pi}{D}\Bigr)\tan\frac{\theta}{2}.\\
\end{dcases}
\end{equation}
This redefinition of the wavevector results in a simpler construction of the reflection matrices $\mathcal{R}_1$ and $\mathcal{R}_2$. As a matter of fact, apart from the conservation of frequency coming from time invariance, the periodicity on the two axes implies that both operators conserve the wavevector $\mathbf{k}$, and that $\mathcal{R}_1$ ($\mathcal{R}_2$) connects different values of $n$ ($m$), while it is diagonal with respect to $m$ ($n$). Finally, it is important to stress two important issues concerning this choice of basis. First, the area of the FBZ is $4\pi^2\sin\theta/D^2$, making the calculation more and more involved for small angles. Moreoever, the redefinition given in Eq.~\eqref{kmn} is not applicable for strictly vanishing $\theta$. In this case we are forced to use the more common definition $\mathbf{k}_n = \mathbf{k}+ 2\pi n/D\,\hat{\mathbf{e}}_x$, and the FBZ simply becomes $-\pi/D<k_x<\pi/D$, whereas $k_y$ takes all real values. We stress at this stage a fundamental difference between the configurations $\theta=0$ and $\theta\neq0$. While for two aligned gratings the $y$ component $k_y$ of the wavevector is strictly conserved in any scattering process, this conservation law is lost even for a small non-vanishing value of the rotation angle $\theta$, since [see Eq.~\eqref{kmn}] changing the value of the diffraction order $m$ modifies the values of both $k_x$ and $k_y$. This breaking of conservation of the component $k_y$ of the wavevector in a 2D-periodic infinite system is the principal reason for the zero-order geometric transition between a 2D- and a 1D-periodic infinite system at $\theta=0$. Finally, we note that in the following the calculation of the reflection matrices for infinite periodic gratings is based on the Fourier Modal Method (FMM)~\cite{Kim12,GranetJOptSocAmA96}. We remark that, while the integration domain with respect to $\mathbf{k}$ in Eq.~\eqref{EqEn} is finite, the number of diffraction orders appearing in Eq.~\eqref{kmn} necessary to cover the entire $\mathbf{k}$ space is in principle unbounded. One thus needs, in any practical numerical implementation of the method, to truncate the Fourier series to a maximum diffraction order $N_\text{max}$, called hereafter truncation order. If both $n$ and $m$ in Eq.~\eqref{kmn} are limited to $N_\text{max}$, the reflection matrices $\mathcal{R}_1$ and
$\mathcal{R}_2$ appearing in Eq.~\eqref{EqEn} have size $2(2N_\text{max}+1)^2\times2(2N_\text{max}+1)^2$. Due to the dependence of the size of the FBZ on the rotation angle $\theta$, $N_\text{max}$ is a function of $\theta$ as well. To give an idea of the truncation order needed in our calculations, we give here two different scenarios: for $\theta=5\,$deg we have chosen $N_\text{max}\simeq60$, while for $\theta=90\,$deg we have taken $N_\text{max}\simeq5$.

The results obtained with the FMM have been verified by means of an independent numerical code, in which a modification to the FMM known as Adaptive Spatial Resolution (ASR) has been implemented. This technique, originally introduced to improve the convergence of the method and to overcome the instabilities observed in particular in the case of metallic gratings~\cite{Granet99,GuizalOpt}, has been recently employed to calculate the radiative heat transfer between two gold gratings~\cite{MessinaPRB17}. For the physical system studied in this work, it has given results in agreement (within the numerical precision) with the ones of the FMM.

\begin{figure}[t!]
\includegraphics[width=0.48\textwidth]{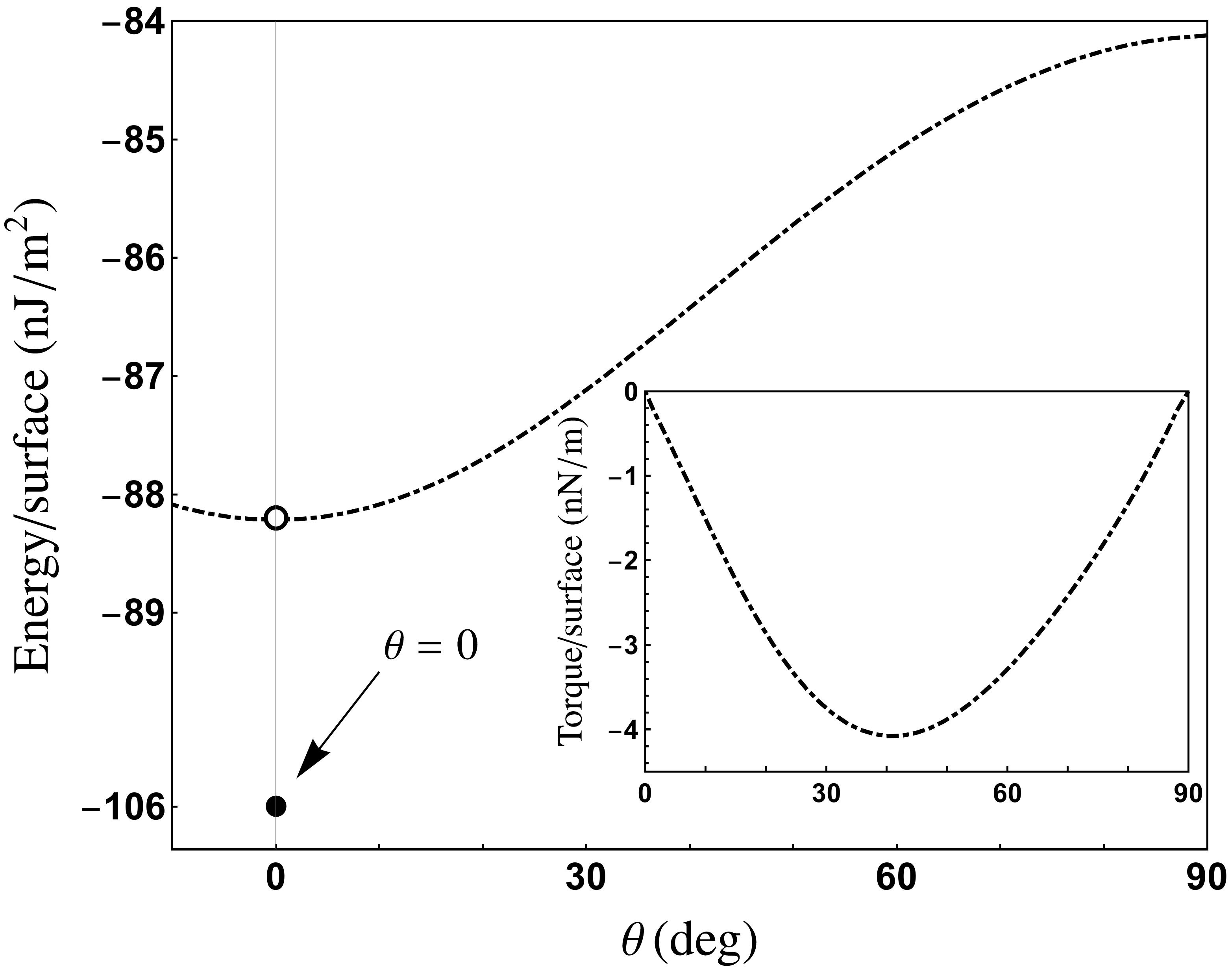}
\caption{Angle-dependent Casimir energy per unit surface between two infinite gold gratings (see text for grating parameters) at distance $d=100\,$nm. The black dot-dashed line corresponds to the energy for all angles $\theta\neq0$, while the black dot shows (not in scale on the vertical axis) the energy between two aligned gratings ($\theta=0$) calculated with a theoretical framework not allowing rotation. The inset shows the torque as a function of $\theta$ for $\theta\neq0$.}
\label{ET_Cas}
\end{figure}

Our first numerical results for this configuration are shown in Fig.~\ref{ET_Cas}, where the energy per unit surface (and the associated torque, in the inset) is plotted as a function of the angle $\theta$ between the gratings. We recognize immediately that the torque tends to zero for the two extreme angles $\theta=0,90\,$deg, an expected result from the symmetry of the system. For intermediate angles, the energy and the torque have a smooth behavior, similar to the one observed in \cite{GueroutEPL15}. For the Casimir energy, it is natural to compare the limit for $\theta$ going to zero (obtained within the theoretical framework adapted to rotated gratings) to the result obtained for perfectly aligned gratings. As shown in Fig.~\ref{ET_Cas}, the latter (the black point, not in scale on the vertical axis) takes on a different value compared to the former, far beyond the numerical error (of the order of 1\% for both results). We will call this jump of the Casimir energy \emph{the $\theta=0$ anomaly}. Because of this discontinuity, the torque is strictly speaking not defined at $\theta=0$ for infinite gratings, while it is defined for any $\theta\neq0$ and tends to zero for $\theta\to0$ .

This intriguing feature has never been pointed out in previous studies of the Casimir torque. In \cite{GueroutEPL15}, based on the assumption of a continuous behavior of the Casimir energy at $\theta=0$, a continuous interpolation function has been used to join the energy at finite $\theta$ with the energy at $\theta=0$ (see Fig.~2 in \cite{GueroutEPL15}), resulting in a Casimir torque which, as we will see, is completely different both qualitatively and quantitatively from what we found. This has dramatic consequences on the Casimir torque between finite-size systems. The reason for the appearance of this anomaly is the breaking of conservation of the $k_y$ component of the wavevector in reciprocal space due to rotation of the system and, as a result, the fundamental change of the structure of reciprocal lattice space.

To explain the $\theta=0$ anomaly, it is instructive to consider the physics produced by the relative rotation of the two gratings. Let us first focus on the scenario of two infinite gratings, rotated by a non-vanishing angle $\theta$. In this case, a given grating line (the axis of a \emph{raised} part of the grating) of grating 2 [say, e.g., the one passing through the origin of the $x-y$ place for any $\theta$, as shown in Fig.~\ref{Geometry}(b)] makes an infinite number of intersections with the gratings lines of grating 1 and, remarkably, this is the case for any value (even extremely small) of $\theta$. This means that, for infinite-size gratings, passing from a finite (albeit small) value of $\theta$ to a strictly $\theta=0$ value implies passing from a configuration with an infinite number of crossing points to a configuration with zero crossing points (the two gratings are perfectly aligned when $\theta=0$). This critical behavior is at the origin of the $\theta=0$ anomaly, and can be interpreted as a transition from a 2D-periodic to a 1D-periodic system. It is absent in all intrinsically anisotropic materials and is a peculiarity of the infinite 1D-grating geometry. Moreover, this jump of the energy at $\theta=0$ no longer occurs in the case of two finite-size gratings. In particular, for two finite identical square gratings [the configuration shown in Fig.~\ref{Geometry}(b)], we can define an angle $\theta_0$ below which no intersection takes place between different grating lines of the two rotated gratings. The configuration drawn in  Fig.~\ref{Geometry}(b) corresponds exactly to $\theta=\theta_0$. In this finite-size scenario the number of crossing points becomes zero for $\theta < \theta_0$, differently from the infinite-grating configuration.
 
We hence expect different behaviors between two finite and two infinite gratings for angles around and below $\theta_0$. In particular:

\begin{enumerate}[(a)]
\item At exactly $\theta=0$, both infinite- and finite-size gratings have no crossing points. Hence if the finite-size gratings are large enough, the Casimir energy will tend to the one for infinite gratings;
\item For angles $\theta\gtrapprox\theta_0$, two sufficiently large finite gratings show a large number of crossing points, hence we also expect that the Casimir energy (and torque) will tend to the one for infinite gratings;
\item For $0< \theta\ll\theta_0$, the behavior of the finite size grating is distinct from the infinite one, in contrast to (a) and (b). Indeed, for finite gratings there are no crossing points between the two gratings, while for infinite gratings we have an infinite number of crossing points. We hence expect that the Casimir energy for finite-size gratings would tend continuously to its value at $\theta=0$, not showing any discontinuous jump at $\theta=0$, contrarily to what happens in the infinite-gratings scenario we already discussed in this section;
\item The position of $\theta_0$ depends on the size of the gratings, and decreases if the grating size increases.
\end{enumerate}

The result of the combination of these four considerations is that, for finite-size systems, in the region $0\leq \theta\lessapprox \theta_0$ the Casimir energy is continuous and the dependence on theta becomes steeper and steeper as the system size increases. This means that the torque per area will exhibit a peak, the height of which increases without bounds when the system size tends to infinity.

\begin{figure}[t!]
\includegraphics[width=0.48\textwidth]{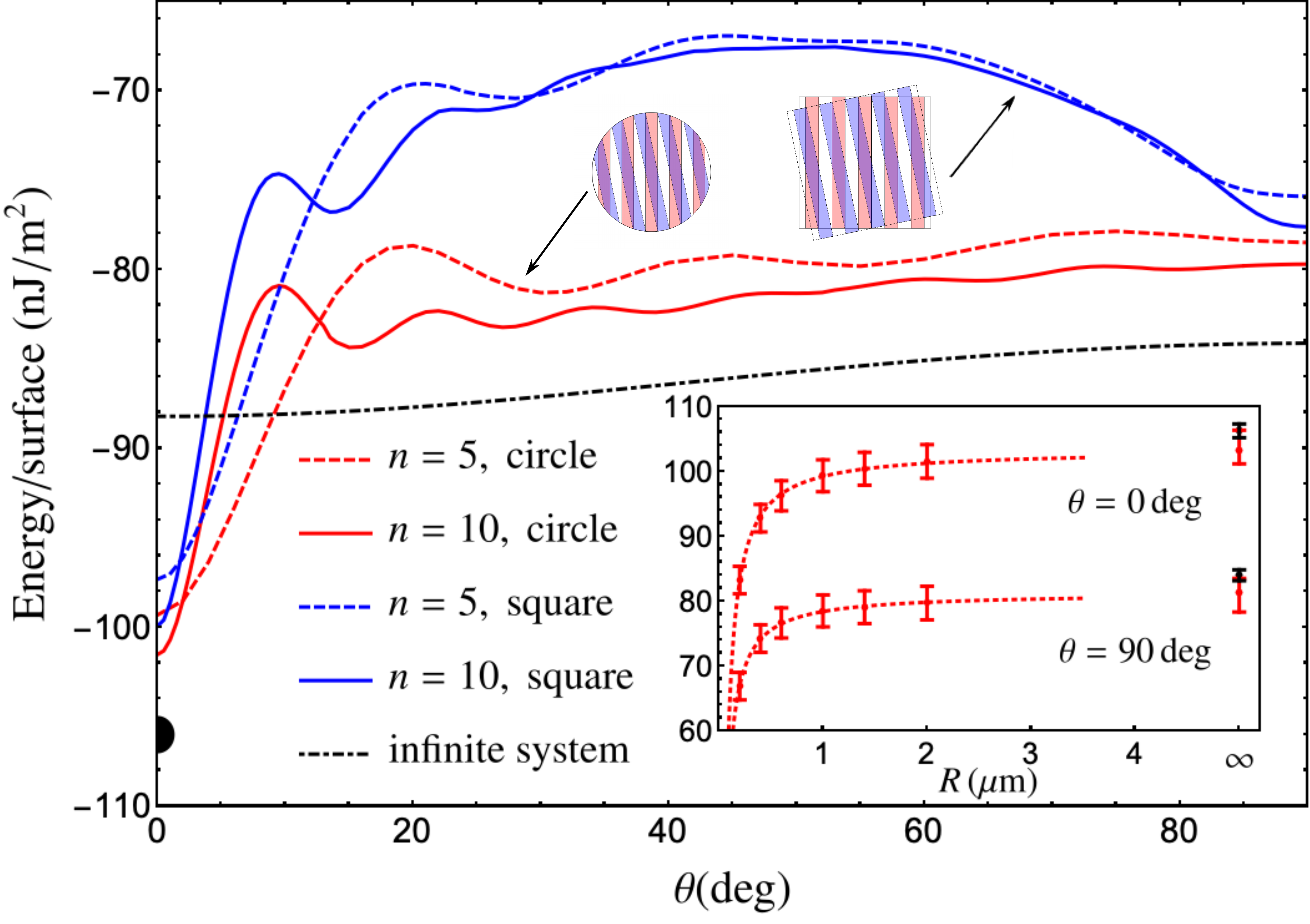}
\caption{Casimir energy per unit surface between two finite gold gratings having $n=5$ (red dashed line for two circular gratings having $R=1\,\mu\text{m}$, blue dashed line for two square gratings having $L=2\,\mu\text{m}$) or $n=10$ (red solid line for two circular gratings having $R=2\,\mu\text{m}$, blue solid line for two square gratings having $L=4\,\mu\text{m}$) unit cells placed at a distance $d=100\,$nm. The black dot-dashed line corresponds to two infinite gratings. We stress that here, unlike the result shown in Fig.~\ref{ET_Cas}, there is no discontinuity in the limit of $\theta$ going to 0. The inset shows the Casimir energy per unit surface (in absolute value) for $\theta=0\,$deg (upper curve) and $\theta=90\,$deg (lower curve) as a function of the radius $R$ of the finite circular grating. The red points obtained numerically are fitted with a function $A+B/R$ (red dotted lines). The asymptotic values for $R\to\infty$ (red dots, not in scale on the horizontal axis) are compared with the ones obtained theoretically for an infinite grating (black dots). All the points are represented with the error bars coming from the respective numerical techniques.}
\label{Energy_finite}
\end{figure}

\begin{figure}[t!]
\includegraphics[width=0.48\textwidth]{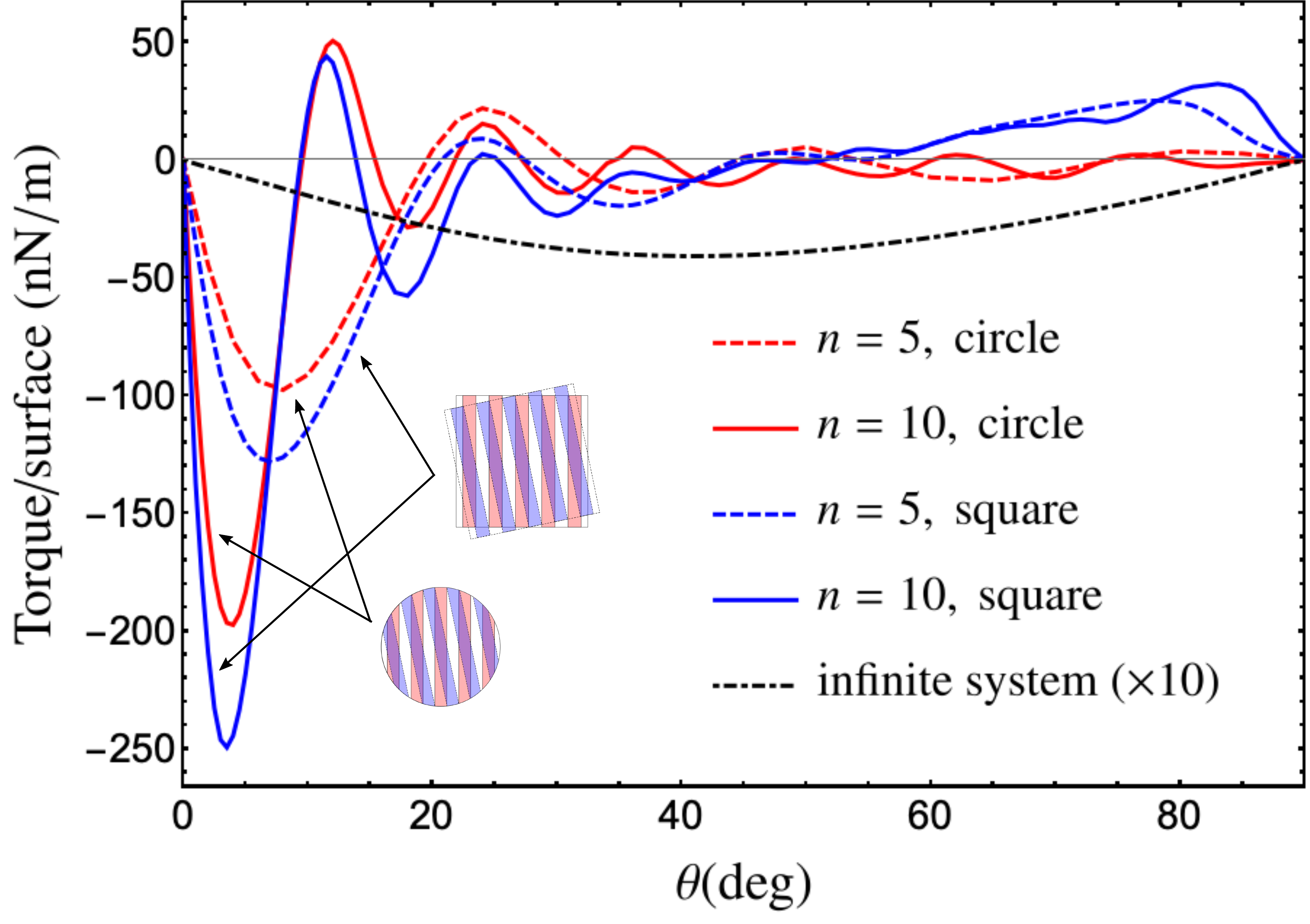}
\caption{Casimir torque per unit surface between two finite gold gratings having $n=5$ (red dashed line for two circular gratings having $R=1\,\mu\text{m}$, blue dashed line for two square gratings having $L=2\,\mu\text{m}$) or $n=10$ (red solid line for two circular gratings having $R=2\,\mu\text{m}$, blue solid line for two square gratings having $L=4\,\mu\text{m}$) unit cells placed at a distance $d=100\,$nm. The black dot-dashed line corresponds to two infinite gratings, multiplied by a factor of 10 in order to make it more visible.}
\label{Torque_finite}
\end{figure}

\subsection{Finite size system: giant Casimir torque}\label{FinSyst}

In order to confirm the predictions we described at the end of the previous section, we focus now on the case of two finite gratings. We consider two different finite-size configurations, namely having circular or square sections, and compute the energy and the torque for different values of the number $n$ of repetitions of the period $D$ along the periodicity axis ($n=5$ in Fig.~\ref{Geometry}). In the case of circular gratings, a given $n$ is associated with a radius $R=nD/2$, while for square gratings it corresponds to a lateral size $L=nD$. The interest of considering the circular geometry is that  its cylindrical symmetry safely ensures that, as in the case of infinite gratings, the absence of the nanostructuring makes the energy angle-independent and thus gives a vanishing torque. In this respect, it is thus more natural to choose this geometry in order to study the limit of infinite size. On the other hand, the corners of the square shape will induce a non-vanishing torque even in the absence of a nanostructuring. This geometry will then provide us with a different interesting insight into geometry-induced finite-size effects. In the square-shape configuration, the analytical expression of the angle $\theta_0$ can be calculated by simple trigonometric arguments. For small $\theta_0$, this can be approximated (in degrees) as $\theta_0(n)\simeq (180/\pi)\arctan(1/n)$, going to zero as expected for increasing $n$.

In the configuration of two finite gratings, numerical computations were performed using {\sc scuff-em}, a free, open-source software implementation of the boundary-element method~\cite{SCUFF1, SCUFF2}. As a first important consistency test, we calculate the Casimir energy for the two extreme angles $\theta=0,90\,$deg for two circular gratings as a function of their radius $R$. The results are shown in the inset of Fig.~\ref{Energy_finite}. In order to extrapolate the results for two infinite gratings, both sets of points have been fitted with a model curve $A+B/R$. The two fitting curves give a good description of the behavior as a function of $R$, and the extrapolated limit $A$ is shown (with associated error bar) in the same curve, not in scale on the horizontal axis. The two values of $A$ (for $\theta=0,90\,$deg) are compared to the results obtained with the previous approach (for infinite gratings), showing that there is good agreement with the code for rotated infinite gratings for $\theta=90\,$deg, and with the code for non-rotated gratings for $\theta=0\,$deg. Apart from providing an important consistency check between the two approaches, this comparison represents a first quantitative confirmation of our physical intuition about the significantly different behavior for small angles between finite and infinite gratings. The main part of Fig.~\ref{Energy_finite} shows the energy per unit surface as a function of the rotation angle $\theta$ for two finite circular or square gratings, having $n=5$ and $n=10$ repetitions, respectively. Several remarks are in order. We first note that, both for circular and square gratings, moving from $n=5$ to $n=10$ repetitions modifies quantitatively and qualitatively the curves. We also observe that for the extreme angles $\theta=0,90\,$deg they move towards the infinite-grating results. Moreover, the $\theta=0$ anomaly is not present in any of these curves, as expected for finite gratings. Furthermore, the comparison between circular and square gratings of the same size directly gives a signature of the role played by geometry in the finite-size scenario. In the case of circular gratings, the energy is increasingly flat for angles close to $\theta=90\,$deg, starting to mimic the behavior of the infinite-grating configuration (black dot-dahsed line in Fig.~\ref{Energy_finite}). On the contrary, even increasing the value of $n$, in the case of square gratings the shape of the energy as a function of the rotation angle is still significantly different from the one associated with two infinite gratings. Finally, concerning the difference between the energy for circular gratings having $n=10$ repetitions and the infinite-grating case, we attribute this disagreement to the relatively small size of the finite systems considered here, that is limited by computation resources.

The torques per unit surface corresponding to the energies analyzed so far are shown in Fig.~\ref{Torque_finite}. In the case of square gratings, we observe that the behavior of the energy, significantly different from the one of two infinite gratings, gives rise to several crossings with the horizontal axis, i.e. the torque changes sign several times between 0 and 90\,deg. This oscillating feature is drastically different from the behavior in Fig.~\ref{ET_Cas}, where the torque is always negative, i.e. always tends to bring the gratings toward the configuration $\theta=0\,$deg. On the contrary, the behavior of the torque between circular gratings is flatter around $\theta=90\,$deg, and shows a much higher agreement with the torque between infinite gratings (note that the latter is multiplied by a factor of 10 in Fig.~\ref{Torque_finite}). We now focus on the most interesting part of these curves, i.e. the region close to $\theta=0\,$deg where, according to our previous analysis, finite-size effects are supposed to be more pronounced. Not only is this behavior very different from the infinite-grating scenario, but the position and the height of the negative peak in the torque is reasonably independent of the geometry. We stress that the first (negative) peak of the torque is also its global maximum (absolute) value in the entire range $[0,90]\,$deg. Figure \ref{Torque_finite} shows us that, when the size of the grating increases, the torque close to $\theta=0\,$deg shows an increasingly deep peak (deviating more and more from the result for infinite gratings), while we expect the behavior on a larger angle region close to 90\,deg to mimic the infinite gratings. We remark that the deep peak in the torque is associated with a slope of the energy as a function of $\theta$ (see Fig.~\ref{Energy_finite}) which becomes steeper and steeper as $n$ increases. It is remarkable that already with $n=10$ period repetitions we observe a maximum torque which is 50 (circular grating) to 60 (square grating) times larger than the one obtained for infinite gratings. The emergence of this maximum of torque for finite-size systems, which we predicted on the basis of the infinite-size calculations (see Sec.~\ref{FinSyst}) is hence confirmed, and is one of the main results of our work.

In order to reinforce our analysis about the connection between this torque enhancement and finite-size effects we go back to our simple geometrical interpretation presented above. We observe that the critical angle $\theta_0$ (below which no intersection between grating lines occurs) equals $\theta_0\simeq11.3\,$deg for $n=5$ and $\theta_0\simeq5.7\,$deg for $n=10$ period repetitions. We now want to show that this angle (whose value is purely based on geometrical arguments) represents a good estimate of the angle at which the large negative peak of the torque occurs. To this aim, we graphically estimate from Fig.~\ref{Torque_finite} the width $\Delta\theta$ of first angular region ($x$-axis) of negative torque including the largest negative peak. The width of this region is almost independent of the grating section (circle or square) and is around $\Delta\theta=21$\,deg for $n=5$ (red and blue dashed lines in Fig.~\ref{Torque_finite}) and $\Delta\theta=9.5$\,deg for $n=10$ (red and blue solid lines in Fig.~\ref{Torque_finite}). If we consider half of these values as a good estimate of the angle associated with the largest negative torque, we clearly see that $\Delta\theta/2\simeq\theta_0$, confirming the validity of our geometrical picture.

Based on this simple model, we are able to provide an estimate of the maximum torque associated to a given grating size (or number $n$ of period repetitions). This estimate coincides with the local maximum in the region $[0,2\theta_0]$ of negative torque. To this end, we start by build a fitting function $f(\theta)$ for the energy associated with a finite grating. Coherently with the vanishing torque, we impose zero derivative at $\theta=0,2\theta_0$. Concerning $f(0)$, we impose the value given by the function $A+B/R$ shown in the inset of Fig.~\ref{Energy_finite}. Finally we approximate $f(2\theta_0)$ with the value of the energy for an infinite grating at the same angle. We stress that this value, as shown in Fig.~\ref{Energy_finite} is lower than the one for a finite grating for $\theta\gtrapprox10\,$deg. As a consequence, we will obtain a conservative value for the maximum torque, lower than the one actually observed for a finite grating. The simplest function able to satisfy these four conditions is a third-order polynomial, from which we estimate the maximum torque as the opposite of the derivative of the fitting function $-f'(\theta_0)$ calculated in the middle of the interval. This procedure gives us an approximate expression $\tau_\text{max}\simeq-56R$ ($R$ expressed in $\mu$m, $\tau$ in nN/m) of the maximum torque, linearly growing as a function of the radius $R$, which is more and more accurate as the $R\to\infty$. Already for small systems with $n=10$ period repetitions, corresponding to $R=2\,\mu$m, we obtain $\tau_\text{max}\simeq-112\,$nN/m, of the same order of the value close to $-200$\,nN/m shown in Fig.~\ref{Torque_finite}.

\section{Conclusions}\label{Conc}

We have shown that the Casimir energy for infinite gratings is discontinuous and displays an anomalous jump at rotation angle $\theta=0$. We explained this behavior in terms of a critical zero-order transition between a 2D-periodic and a 1D-periodic system, and showed that this gives rise to a Casimir torque which is both qualitatively and quantitatively different from what was previously predicted. On the basis of these unexpected findings we studied the Casimir energy and torque for finite-size systems. We showed that they exhibit new and strikingly different features: several sign changes and a giant torque per unit area at small angles, whose amplitude remarkably increases without bounds with the size $R$ of the system, and in particular linearly with $R$ for large system sizes. We showed that for finite gratings with only $n=10$ period repetitions, the maximum torque is already 50-60 times larger than the one predicted in the case of infinite gratings. We explained this giant torque and studied its tunability. These findings offer a comprehensive study of the Casimir torque with respect to system size. The giant Casimir torque paves the way to the possibility of an experimental measurement of the rotational effects induced by quantum fluctuations across a vacuum gap and to new opportunities to exploit the vacuum field to realize a contactless quantum vacuum torsional spring, with promising applications in micro- and nano-technological systems and devices.

\begin{acknowledgments}
M.~A. thanks the support from Institute Universitaire de France. Research by M.~A., B.~G. and R.~M. was carried out using computational resources of the group Theory of Light-Matter and Quantum Phenomena of the Laboratoire Charles Coulomb. V.~N.~M. thanks the group Theory of Light-Matter and Quantum Phenomena of the Laboratoire Charles Coulomb for invitation to Montpellier and financing the visit. V.~N.~M. acknowledges St. Petersburg State University for a research grant, Grant No. IAS\_11.40.538.2017. Research by V.~N.~M. was performed at the Research park of St. Petersburg State University ``Computing Center". H.~B.~C. and M.~W. are supported by HKUST 16300414 and AoE/P-02/12 from the Research Grants Council of Hong Kong SAR
\end{acknowledgments}

\end{document}